\documentclass[10pt,journal,cspaper,compsoc]{IEEEtran}
\ifCLASSOPTIONcompsoc
   \usepackage[nocompress]{cite}
\else
   \usepackage{cite}
\fi
\ifCLASSINFOpdf
  \usepackage[pdftex]{graphicx}
\else
\usepackage{graphicx}
\fi
\usepackage[cmex10]{amsmath}
\usepackage{amssymb}
\usepackage{graphicx}
\usepackage{mathpartir}
\usepackage{amssymb}
\usepackage{amsfonts}
\usepackage{stmaryrd}
\usepackage{rotating}
\usepackage{listings}
\usepackage{xcolor}

\newcommand{\Class}[0]{\ensuremath{\textbf{Class~}}}
\newcommand{\feature}[0]{\ensuremath{\textbf{feature~}}}
\newcommand{\enD}[0]{\ensuremath{\textbf{end~}}}

\newcommand{\translate}[1]{\ensuremath{\llbracket#1\rrbracket}}
\newcommand{\shout}[1]{\mathcal{W}\translate{#1}}
\newcommand{\inputVal}[0]{\mathcal{I}\translate{\cdot}}
\newcommand{\attached}[0]{\ensuremath{\textbf{attached~}}}
\newcommand{\detachable}[0]{\ensuremath{\textbf{detachable~}}}

\usepackage{array}
\usepackage{multirow}
\usepackage{url}
\usepackage{hyperref}
\usepackage{xspace}
\usepackage{dsfont}

\begin{document}
%
\title{Class Schema Evolution for Persistent Object-Oriented Software: Model, Empirical Study, and Automated Support}
%
%
%
%

\author{Marco~Piccioni,~\IEEEmembership{Member,~IEEE,}
        Manuel Oriol, and~Bertrand~Meyer
\IEEEcompsocitemizethanks{\IEEEcompsocthanksitem Marco Piccioni and Bertrand Meyer are with the Chair of Software Engineering, ETH Zurich, Switzerland.\protect\\
E-mail: marco.piccioni@ieee.org, Bertrand.Meyer@inf.ethz.ch
\IEEEcompsocthanksitem Manuel Oriol is with ABB Corporate Research,  Industrial Software Systems, Baden-D\"attwil, Switzerland and the University of York, UK. \protect\\
E-mail: manuel.oriol@ch.abb.com, manuel@cs.york.ac.uk}
\thanks{}}

\markboth{IEEE Transactions on Software Engineering}%
{}

%


\IEEEcompsoctitleabstractindextext{%
\begin{abstract}
With the wide support for object serialization in object-oriented programming languages, persistent objects have become common place and most large object-oriented software systems rely on extensive amounts of persistent data. Such systems also evolve over time. Retrieving previously persisted objects from classes whose schema has changed is however difficult, and may lead to invalidating the consistency of the application. 

The ESCHER framework addresses these issues through an IDE-integrated approach that handles class schema evolution by managing versions of the code and generating transformation functions automatically. The infrastructure also enforces class invariants to prevent the introduction of potentially corrupt objects. This article describes a model for class attribute changes, a measure for class evolution robustness, four empirical studies, and the design and implementation of the ESCHER system.
\end{abstract} 

}

\maketitle

\IEEEdisplaynotcompsoctitleabstractindextext

%
\IEEEpeerreviewmaketitle

\section{Introduction}\label{sec:intro}

\IEEEPARstart{B}{etween} the time an object-oriented program writes objects to persistent storage and when another execution retrieves them, the program may have evolved; in particular, classes describing these objects may have changed. The need to fit old objects into new classes is the problem of schema evolution for object-oriented software. While many solutions have been suggested and are currently used, none has been to our knowledge generally accepted.

The techniques used to cope in practice rely heavily on manual effort by the developers, who must examine previous class versions and provide conversion code. This approach is not only tedious but a threat to software reliability, as it usually relies on tolerant retrieval algorithms that make questionable decisions about the key issue: how to avoid accepting semantically inconsistent objects into the retrieving system. 

The purpose of the present work is to lay a solid foundation for a general, stable solution to the problem of object-oriented class schema evolution. The contribution includes four major components: empirical studies; a formal model, resulting both from the empirical studies and from object-oriented theory; a measure of persistence evolution robustness and a schema evolution tool implementing the model.

Section~\ref{sec:motivation} presents a motivating example and the corresponding ESCHER tool support. Section~\ref{sec:earlierWork} describes previous approaches to class schema evolution. Section~\ref{sec:model} defines a formal model representing atomic attribute changes through a set of transformation functions, and suggests a definition of a measure for evolution robustness. The empirical studies in Section~\ref{sec:dataAnalysis} are essential to understand what kind of schema evolution actually happens in practice; without empirical evidence, proposed solutions to the schema evolution problem are bound to be off the mark. The first study analyzes the evolution of serializable classes in five versions of the \texttt{java.util} package. 
It highlights the acuteness of the problem by showing that 17.7\% of the changes affect the persistence of the serializable classes, and justifies the set of atomic attribute changes considered in the model. The measure for evolution robustness is also calculated to provide further evidence. The second study covers, for balance, the history of five versions of the Eiffel base library classes.
The findings confirm those of the first study and lead to a refined set of atomic attribute changes. Two more studies analyzing persistent classes in two well known software projects, the Apache Tomcat web server and the EiffelStudio IDE (development environment), lead to similar conclusions. The implementation (Section~\ref{sec:implementation}) is the ESCHER tool, seamlessly integrated into EiffelStudio and supporting developers through the entire schema evolution process, from class release time to retrieval time. It includes version handling, code template generation for transformation functions and a retrieval algorithm preventing the acceptance of inconsistent objects into the system. Section~\ref{sec:assessment_future_work} evaluates ESCHER by analyzing which code changes ESCHER actually detects in the four projects mentioned above. It then analyzes the advantages and limitations of the contribution, drawing conclusions and suggesting future work. This paper extends previous work from the authors \cite{PiccioniOriolMeyer09} by adding three new studies (on the EiffelBase library, on the Java Apache Tomcat repository and on the EiffelStudio IDE repository), by updating the model, by defining and applying a measure for evolution robustness to all the classes in the \texttt{java.util} package, and by refining the tool implementation.
\section{A Schema Evolution Example and its Handling in ESCHER}\label{sec:motivation}
This section presents an example scenario and shows how the ESCHER tool copes with it. 
\subsection{A case of object-oriented schema evolution}
Assume there is a software system storing objects of type \textit{BANK\_ACCOUNT} (see Figure~\ref{fig:ver1}). 
\begin{figure}[ht]
\begin{lstlisting}
class
	BANK_ACCOUNT
create make

feature 	-- Initialization
	make
			-- Create a bank account with an initial deposit.
		do
			tot_deposits := 1
		end	
feature 	-- Status report
	balance: INTEGER
			-- Account balance.
		 do
		 	Result := tot_deposits - tot_withdrawals
		 end		
	info: STRING
			-- Some numeric information.		 
feature	-- Basic operations
	deposit (sum: INTEGER)
			-- Add `sum' to account.
		require
			sum_non_negative: sum > 0
		do
			tot_deposits :=  tot_deposits + sum
		ensure
			balance_correct: balance = old balance + sum
 		end	

	withdraw (sum: INTEGER)
			-- Retrieve `sum' from account.
		require
			sum_non_negative: sum > 0
			has_sufficient_funds: sum < balance
		do
			tot_withdrawals :=  tot_withdrawals + sum
		ensure
			balance_correct: balance = old balance - sum
 		end
feature {NONE} -- Implementation
	tot_deposits: INTEGER
			-- Total amount deposited.	
	tot_withdrawals: INTEGER
			-- Total amount withdrawn.
invariant
	valid_account: tot_deposits > tot_withdrawals
end
\end{lstlisting}
\caption{\textit{BANK\_ACCOUNT}, version 1.}\label{fig:ver1}
\end{figure}
The balance is computed on-demand from the total amount of deposits and withdrawals, and there is an explicit class invariant capturing the bank account intended semantics. This invariant is checked after the invocation of the constructor \textit{make} of a bank account object, and also before and after the invocation of any other routine in the class.

\begin{figure}[ht]
\begin{lstlisting}
class
	BANK_ACCOUNT
create make

feature 	-- Initialization
	make
			-- Create a bank account with an initial deposit.
		do
			balance := 1
		end
feature 	-- Status report
	balance: INTEGER
			-- Account balance.
	info: INTEGER
			-- Some numeric information.		
feature	-- Basic operations
	deposit (sum: INTEGER)
			-- Add `sum' to account.
		require
			sum_non_negative: sum > 0
		do
			balance :=  balance + sum
		ensure
			balance_correct: balance = old balance + sum
 		end

	withdraw (sum: INTEGER)
			-- Retrieve `sum' from account.
		require
			sum_non_negative: sum > 0
			has_sufficient_funds: sum < balance
		do
			balance :=  balance - sum
		ensure
			balance_correct: balance = old balance - sum
 		end
invariant
	valid_account: balance > 0
end
\end{lstlisting}
\caption{\textit{BANK\_ACCOUNT}, version 2.}\label{fig:ver2}
\end{figure}
Version 1 then evolves into version 2 (Figure~\ref{fig:ver2}), in which the query \textit{balance} becomes an attribute updated every time a deposit or a withdraw takes place, and the attribute \textit{info} becomes an integer. While in languages with a C-derived syntax like Java we would need to modify client code to accommodate the fact that a method becomes an attribute, here clients are not affected, because in Eiffel both attributes and argument-less functions can be accessed in the same way from outside the class, providing uniform access to the class itself. Note also how the invariant is now expressed in terms of the new attribute \textit{balance}.
What happens when trying to retrieve an object stored with version 1 of class \textit{BANK\_ACCOUNT} into an object of version 2 depends on the retrieval system. Silently accepting objects having the attribute \textit{balance} initialized to the default is in this case wrong. To understand why, imagine developers forget to code an appropriate transformation function assigning the difference between \textit{tot\_deposits} and  \textit{tot\_withdrawals} stored in version 1 to the newly created \textit{balance} attribute. Then \textit{balance} will be zero for all retrieved objects, which is hardly ideal. If we consider the Eiffel version 2 of class \textit{BANK\_ACCOUNT} instead, at deserialization time there will be an invariant failure triggering an exception, because the value of \textit{balance} is not positive as the class invariant prescribes. 

\subsection{The ESCHER solution}
ESCHER aims at supporting developers in evolving object-oriented applications. It integrates seamlessly into an IDE, keeping track of all class versions and system releases. When a coherent set of classes is released, developers are supported in writing transformation functions to migrate from one version to another. Depending on the specific evolutionary changes automatically detected, ESCHER support can vary from a function template containing some standard initializations to a full-blown function body generation, where no additional developer intervention is needed. Focusing on the previous example, we now show a worst-case and a best-case scenario with respect to code generation. For the \textit{balance} attribute, the tool recognizes that two attributes have disappeared and a new one has been added, but obviously needs the developer's help to figure out how to initialize it. On the bright side, a default value initialization will raise an exception because it violates the version 2 class invariant, so there is no risk of accepting inconsistent objects into the system. For the \textit{info} attribute, ESCHER recognizes the type change and provides a full code generation of the transformation function body. Finally, ESCHER includes a deserialization algorithm that is aware of the presence (or absence) of transformation functions between any two versions of each class, triggering deserialization failures if the required transformation function is not present, or if the deserialized object does not satisfy the new version class invariant. 
\section{Previous Approaches}\label{sec:earlierWork}
The issues arising from schema evolution are widely acknowledged. They affect object-oriented databases, relational databases and all programming languages providing a serialization facility. 
\subsection{The two-version approach}
Given a class whose objects we need to serialize, the most widespread approach takes into account the ``current'' version and the ``stored'' one. To convert objects between the two, developers typically implement a transformation function that should be able to handle all possible stored versions of the data. 

In Java a class can enable future serialization of its instances by implementing the \texttt{Serializable} interface, or its descendant \texttt{Externalizable}~\cite{Atkinson00,Jordan04, javaspecs}. Developers can provide custom deserialization methods to help establishing the class invariant. 

The .NET framework, in a way similar to Java and starting from version 2.0, provides Version Tolerant Serialization (VTS)~\cite{vts}. In spite of the name, VTS does not implement true version support. The serialized object and all the referenced objects in the object graph have to be tagged with the \texttt{Serializable attribute}. By using ``attributes'' instead of interfaces the mechanism is decoupled from the class hierarchy. As in Java, object constructors are not invoked during the deserialization process, so implicit class invariant violations may happen if developers do not provide custom deserialization methods. 

Eiffel's serialization mechanism also offers a solution in which all conflicts are resolved in one class. Custom deserialization behavior is available by inheriting from a \texttt{MISMATCH\_CORRECTOR} class and redefining a callback to re-establish the class invariant. As a difference with the solutions above, it is easier to detect class invariant violations because of the native support for Design by Contract~\cite{Meyer92}.

In our view, the serialization mechanisms of O-O languages are converging towards a full-fledged solution such as an object oriented database management system (OODBMS). To assess such a solution, we examine the db4o object oriented database~\cite{PatersonEdlich06}.\footnote{http://developer.db4o.com/Home.aspx (last visited: 15/6/12)}
Db4o stores objects as they are, without polluting them with persistence code. With respect to schema evolution handling, if developers need a custom behavior to establish the retrieving object class invariant, they have two possibilities: choose to use reflectively invoked methods in the object class or register listeners to specific \texttt{ObjectContainer} events outside the object class. As seen for Java and .NET, there is not explicit support for class invariant violation. 

The Orion object-oriented database~\cite{BanerjeeKimKimKorth87} introduces a model recognizing the importance of invariants to validate conversions. Compared to ESCHER, the Orion model is more complex because it deals with a broader set of code changes, mostly involving single and multiple inheritance, while ESCHER considers the object's flattened form. Though simpler, the ESCHER model is in one case more general, because it allows switching to any type when detecting an attribute type change, while Orion only allows switching to an ancestor type. ESCHER also offers a limited support for ``attribute renamed'', while Orion does not take any action in this case. Finally, in Orion invariants are mostly related to keeping the system consistent with respect to inheritance, while ESCHER uses class invariants to validate the conversions semantically. 

\subsection{The multi-version approach}
By storing all versions of a class, the multi-version approach makes it easier to provide more specific schema evolution support. ENCORE~\cite{SkarraZdonik86} is an early example enforcing a serialized form limited to the attributes in the class interface, and proposing an ad hoc constraint language for inter-property constraints.

A general framework similar to ESCHER and applied to an OODB was proposed by Lautemann~\cite{Lautemann96}. It required a schema designer role conceptually distinct from the application developer role, and envisioned a completely transparent schema evolution that we believe cannot be achieved. 

A more recent example of a multi-version OODB is Versant Fastobjects.\footnote{www.versant.com (last visited: 15/6/12)} It consists of two main storage areas: a dictionary, which carries the class definitions for objects in the database, and the actual database, containing the serialized objects forms. The dictionary is necessary to understand the structure of the objects stored in the database. The delicate issue consists in modifying the dictionary after having stored some objects. The dictionary keeps versions for every class, so when a class schema modification is detected, old objects are ``transparently'', lazily, and possibly inconsistently converted. 

To convert a stored object of a certain version into an object of the current version, CLOSQL uses update or backdate routines~\cite{MonkSommerville93}. An unfortunate requirement is that a database administrator is needed every time a class is created, to specify which update or backdate routines have to be executed. As in our approach, the update and backdate routines for a certain version are kept all together, in this case in an ``update method''. CLOSQL supports ``linear versioning'': a new version can only be generated from the latest version. In contrast to CLOSQL, we handle transformation functions also from older versions with respect to the last one.

GemStone is a computationally complete Smalltalk extension providing data definition, data manipulation and query facilities to persist objects across executions~\cite{GemStone09}. Each class in GemStone has an associated class history containing all the previous versions. Developers decide when to define a new version and when to override ad hoc migration methods in the destination class.    

The Oracle relational database, starting from version 11g release 2, has introduced ``editions'' to offer better support for online application upgrade~\cite{OracleEditions09}. The concept is more restrictive with respect to what the authors call releases, because it is limited to ``editionable'' object types, which do not include, for example, tables and java classes.   
 
When a certain class evolves over time, it may be considered a different type, and named differently, or it may be considered the same type, keeping the same name but providing other means of taking into account the different inner structure and semantics. While the second option is mainstream, the first one has been previously explored by the authors~\cite{PiccioniOriolMeyer07}. Every new version of a class is given a new name including version information. Versions can therefore be considered full citizens of the type system. To cope with this scenario, Eiffel provides \texttt{converters}~\cite{Meyer01}. Converters are a language mechanism intended to cover those cases in which inheritance is not appropriate, for example to convert string implementations across different systems. A downside of this approach is the lack of scalability. In fact every type should provide converters for each previous version, therefore clogging the corresponding class code with a potentially high number of transformation functions.

Another example of making different, parallel versions explicit can be found in UpgradeJ~\cite{BiermanParkinsonNoble08}, though focused neither on object persistence nor on enforcing semantical consistency. 

The automated detection of code changes has been extensively explored in literature, both with respect to existing libraries and software configuration systems, for example by Dig \textit{et al.}~\cite{DigManzoorJohnsonNguyen07}. One interesting approach proposes that schema evolution can be solved by devising bi-directional transformations ~\cite{CzarneckiFosterHuLammelSchurrTerwilliger09,HofmannPierceWagner10}. Though acknowledging the advantages of generating the inverse transformations (mostly) for free from a bi-directional domain specific language, we believe that using it would be an excessive burden on the developer. We prefer to keep one programming language and provide  transformation functions with explicit contracts to improve their reliability.  

\subsection{Other approaches}\label{sec: ourApproach}
Orthogonal Persistence Java (OPJ) is an extension of the Java Language Specification adding ``orthogonal persistence'' capabilities to the Java platform~\cite{Atkinson78,AtkinsonMorrison95}. The well-known prototype implementation for OPJ, PJama~\cite{AtkinsonJordan00,LewisMathiskeGafter00,Dmitriev01} is an extension of the Java Virtual machine together with a persistent store, in which the state of an executing application is kept. The system state is check-pointed atomically and periodically to be able to recover from exceptions and crashes. PJama provides an approach to schema evolution that involves persisting both objects and classes. To perform conversions between classes, PJama developers use a small API and a standalone, command-line utility. In case the changes are validated, objects are typically converted eagerly.

Several authors suggest to assist metamodel evolution by stepwise adaptation in a similar fashion with respect to what is presented in this paper, using a transformational approach and a classification of modifications~\cite{Wachsmuth07,GruschkoKolovosPaige07,HerrmannsdoerferBenzJuergens08,CicchettiDiRuscioEramoPierantonio08}. Analyzing the AST in metamodels and applying similarity metrics to detect changes has been done by Falleri \textit{et al.}~\cite{FalleriHuchardLafourcadeNebut08}. All the cited works about metamodels are of interest because class schemas are a particular kind of metamodels. 

The automatic generation of transformation functions through type transformers in C programs has also been described by Neamtiu \textit{et al.}~\cite{NeamtiuHicksStoyleOriol06}. It focuses on updating structs whose layout might evolve. However,  it is not per se linked to object-orientation and it does not benefit from having a model or an integration into an IDE. 

The formal object model proposed by the ODMG standard~\cite{ODMG00} encompasses more definitions than ESCHER's, but it is missing some modifications on attributes that we have seen being relevant in practice, like renaming and changing type~\cite{DelgadoSamosTorres03}.

Schema evolution issues are also relevant for relational databases~\cite{Date03}. The inference rules we derive in Section~\ref{sec:model} are related to those implemented in many schema matching tools available~\cite{RahmBernstein01}. In particular, TransScm and Tess~\cite{MiloZohar98,Lerner00} view a schema as a set of types and provide two-version transformations following certain rules. Human intervention is required to provide new rules or to select match candidates. Apart from the different transformation functions semantics and implementation, the most important differences are that ESCHER keeps track of all versions of a class schema, and limits the conversions to those for which our data surveys show potential application. 

The PRISM workbench~\cite{CurinoMoonZaniolo08,SchemaEvolutionBenchmark} presents some analogies with ESCHER, namely the use of a GUI-based tool to facilitate schema evolution, and a certain degree of automation in the process. The main difference we noticed is that in PRISM the database administrator has to input manually the Schema Modification Operators (SMO), while in our case they are detected automatically while the system keeps track of all versions. An additional difference is that the SMOs, though more numerous than our atomic attribute changes, do not include ``column type change'', an analog of our ``attribute type changed''.
\section{The Escher Model of Object-Oriented Software Evolution}\label{sec:model}

This section presents a model of updates that particularly emphasizes the generation of transformation functions --- contrary to previous models 
\cite{BoyapatiLiskovShriraMohRichman03, FranconiGrandiMandreoli01}.

\begin{figure*}[!ht]
{\small
\[
\begin{array}{c}
\begin{array}{l c l r}
\multicolumn{3}{l}{C,D \in Names} & class~names\\
\multicolumn{3}{l}{N,M \in GNames} & names~for~generic~parameters\\
\multicolumn{3}{l}{name \in names} & names~for~attributes\\
att &::=&name: type& attribute~definition\\
&|&name: N& attribute~of~generic~type\\
type &::= & C& class~type\\
&|&type[type]&generic~derivation\\
&|&\attached C& void-safe~type\\
&|&\detachable C& non~void-safe~type\\
class & ::= & \Class C~  \feature att_1,\ldots, att_n~\enD & class\\
&|& \Class C [N_1,\ldots,N_i]~\feature att_1,\ldots,att_n~\enD &generic~class\\
\end{array}\\
\\
\hline
\\
\begin{array}{c}
{att_{i} \in class_0};~A_{noChange(att_{i})}(class_0)=class_0\\
\\
{att \notin \left\{{att_{1},\ldots,att_{n}}\right\}};~A_{new(att)}(\Class C\ldots  \feature att_1,\ldots, att_n~\enD)=\\
\Class C\ldots  \feature att_1,\ldots, att_n, att~\enD\\
\\
\inferrule{\begin{array}{c}
(att_i=name:N \wedge att'_i=name':N)\vee(att_i=name:type \wedge att'_i=name':type)\\
{att_{i} \in class_0};~class_0=\Class C\ldots \feature att_0,\ldots,att_n~\enD\\
class_1=\Class C\ldots \feature att_0,\ldots,att_{i-1},att'_i, att_{i+1},\ldots,att_n~\enD
\end{array}}
{A_{nameChange(name,name')}(class_0)=class_1}\\
\\
\inferrule{\begin{array}{c}
{type\neq type'};~att_i=name:type \wedge att'_i=name:type'\\
{att_{i} \in class_0};~class_0=\Class C\ldots \feature att_0,\ldots,att_n~\enD\\
class_1=\Class C\ldots \feature att_0,\ldots,att_{i-1},att'_i, att_{i+1},\ldots,att_n~\enD
\end{array}}
{A_{typeChange(name:type,name:type')}(class_0)=class_1}\\
\\
\inferrule{\begin{array}{c}
att_i=name:type\\
{att_{i} \in class_0};~class_0=\Class C\ldots \feature att_0,\ldots,att_n~\enD\\
class_1=\Class C\ldots \feature att_0,\ldots,att_{i-1}, att_{i+1},\ldots,att_n~\enD\\
\end{array}}
{A_{removeAttribute(name)}(class_0)=class_1}\\
\\
\inferrule{\begin{array}{c}
att_i=name:C\\
{att_{i} \in class_0};~class_0=\Class C\ldots \feature att_0,\ldots,att_n~\enD\\
class_1=\Class C\ldots \feature att_0,\ldots,att_{i-1}, name:\attached C,att_{i+1},\ldots,att_n~\enD\\
\end{array}}
{A_{addAttach(att_{i})}(class_0)=class_1}\\
\end{array}
\end{array}
\]
}
\caption{Classes and SMOs.}\label{fig:classdef}
\end{figure*}
The upper part of Figure~\ref{fig:classdef} presents a simplified syntax of class definitions in Eiffel
programs. We omit the declaration both of routines and constraints on generic parameters as they are typically not included in the serialized form. We also do not explicitly consider inheritance, both single and multiple, because we have access to the flattened class schema. This is a valid assumption as serialized objects are de facto flattened as well. 

The following subsections define atomic attribute changes and class transformations, how to
generate a type transformer from a class transformation, and present strategies to extract type transformers.

\subsection{Classifying software evolution: SMOs}

Adapting the definition from Curino \textit{et al.}~\cite{CurinoMoonZaniolo08,SchemaEvolutionBenchmark} to an object-oriented context, we call Schema Modification Operator (SMO) an incremental, atomic change to a class schema, along with the semantic changes possibly occurring to any instances conforming to that schema. More formally, a SMO is a function modifying at most one attribute: \[
A: class\mapsto class
\]

We define six standard SMOs: 

\begin{itemize}
\item Attribute not changed
\item Attribute added
\item Attribute renamed
\item Attribute type changed
\item Attribute removed
\item Attribute attached added
\end{itemize}

 The bottom part of Figure~\ref{fig:classdef} shows the definition of the six kinds of SMOs.

A \textit{class transformation} $T_{A_1,\ldots,A_N}$ going from one version of a class to 
another can then be described by a list of SMOs ${A_1,\ldots,A_N}$
($N\geq 1$): 

\[
\begin{array}{l}
T_{A_1,\ldots,A_N}: class\mapsto class\\
\end{array}
\]
such that:
\[
\begin{array}{l}
T_{A_1,\ldots,A_N}(class_0)={(A_n\circ\ldots\circ A_1)}(class_0)
\end{array}
\]

Note that any modifications to the attributes of a class can be described by some class transformation --- for example by deleting all attributes whose names are not present in the class anymore and adding the new ones. Thus, class transformations are complete with respect to attribute modifications. This can be seen easily: suppose we have two classes C1 and C2 with n attributes and m attributes respectively, and assume the generic parameter list is preserved between C1 and C2. Therefore there always exists a class transformation from C1 to C2 that first deletes all the n attributes from C1 and then adds all the m attributes to C2. 
While there is always a decomposition, using a straightforward algorithm might not produce the best results, so devising a heuristic is essential to use this model as the basis of an implementation.

\begin{figure*}[!ht]
\begin{center}
{\[\small
\begin{array}{c}
\begin{array}{l c l r}
\multicolumn{3}{l}{otname \in OTNames} & object~transformers~names\\
e&::=& e.e~|~\textbf{oldc}~|~name&expressions\\
instruction &::=&\textbf{Result.}name\textbf{:=}e~|~\textbf{noop}~|~otname(e,e)& instructions\\
instructions &::=&instruction~|~instruction;instructions&list~of~instructions\\ 
ot&::=& otname(oldc: oldC): C~instructions&object~transformer\\
\end{array}\\
\\
\hline
\\
\begin{array}{l c l}
\multicolumn{3}{l}{\mathcal{G}\translate{\cdot}:~(class\mapsto class)\mapsto ot}\\
\multicolumn{3}{l}{\mathcal{G}\translate{T_{A_1,\ldots,A_n}}=otname(oldc: oldC):C~\mathcal{S}\translate{T_{A_1,\ldots,A_N}}}\\
\\
\multicolumn{3}{l}{\mathcal{S}\translate{\cdot}:~(class\mapsto class)\mapsto instructions}\\
\mathcal{S}\translate{A_{noChange(att_{i})}}&=&\textbf{Result.}name_{i}\textbf{:=}\textbf{oldc.}name_i\\
\mathcal{S}\translate{A_{new(att)}}&=&\textbf{Result.}name\textbf{:=}\inputVal\\
&&where~att=name:\ldots\\
\mathcal{S}\translate{A_{nameChange(name:type,name':\textsl{type})}}&=&\textbf{Result.}name'\textbf{:=}\textbf{oldc.}name\\
\mathcal{S}\translate{A_{typeChange(name:type,name:\textsl{type'})}}&=&\textbf{Result.}name\textbf{:=}\textbf{oldc.}name\\
&&if~\textsl{type'}~is~assignable~to~\textsl{type}\\
&=&\textbf{Result.}name\textbf{:=}otname(\textbf{oldc.}name)\\
&&if~there~is~an~object~transformer~otname\\
&&transforming~an~object~of~type~\textsl{type}~\\
&&into~an~object~of~type~\textsl{type'}\\
&=&\shout{name_0}~otherwise\\
\mathcal{S}\translate{A_{removeAttribute(name_0)}}&=&\shout{name_0}\\
\mathcal{S}\translate{A_{addAttach(att_{i})}}&=&\textbf{Result.}name_{i}\textbf{:=}\textbf{oldc.}name_i\\
\mathcal{S}\translate{T_{A_1,\ldots,A_n}}&=&\mathcal{S}\translate{A_1};\mathcal{S}\translate{T_{A_2,\ldots,A_N}}\\
\end{array}
\end{array}
\]}
\caption{Object transformers.}\label{fig:ttsyn}
\end{center}
\end{figure*}

\subsection{Object transformers}
While SMOs allow an easy representation of the static transformations of a class, this is not sufficient to generate the object transformers that create an instance of the new class from a serialized instance of the old class. The main reason is that there is a need for an explicit default initialization of new attributes. To define such initializations, programmers actually need to input default initialization values. To represent that, we use the token $\inputVal$ (for input), evaluating into the next value in a list of inputs. We also add an output generator $\shout{\cdot}$ (for warning), warning programmers of the impossibility of generating a translation or a removal operation and evaluating into ``noop''. The upper part of Figure~\ref{fig:ttsyn} presents a succinct syntax for object transformers. The generation of object transformers from a class transformation can be expressed as transformation functions as shown in the bottom part of Figure~\ref{fig:ttsyn}. The function $\mathcal{G}\translate{\cdot}$ takes a class transformation and generates the associated object transformer. Finally, the function $\mathcal{S}\translate{\cdot}$ takes a class transformation and generates the actual object transformer code.

\subsection{Heuristics for class schema evolution}
While we are currently considering the extraction of transformations based on actual changes made by programmers using the IDE, the current solution relies on statically comparing the abstract syntax trees (ASTs) to detect SMOs. Static comparisons imply that in certain situations, like the ``Attribute renamed'' SMO below, the outcome of the tool can only suggest the possibility that the 
specific SMO has happened. The extraction relies on a set of heuristics successively applied:
\begin{enumerate}
  \item An attribute that does not change name and declared type generates an ``Attribute not changed'' SMO. A possible change in the attribute semantics will be taken care by the retrieving object class invariant.
  \item An attribute that does not change name but changes type between two versions generates an ``Attribute type changed'' SMO. The framework is always able to detect this SMO, and depending on the type involved, to provide a complete code generation (as in the example in Section~\ref{sec:motivation}). A change in the attribute semantics is taken into account by the new type and by the retrieving object class invariant.
  \item An attribute \textit{att} removed from the old version is a candidate for generating an ``Attribute renamed'' SMO if in the new version there is a new attribute \textit{att'} having the same type as \textit{att}. By only comparing the two classes' ASTs it is not possible to determine if a rename actually occurred, because the new attribute could be semantically unrelated to the old one. In this case we would have an "Attribute removed" followed by an "Attribute added". The framework detects that this SMO might have occurred and generates a transformation function template warning about the possibility of a rename. In this case the retrieving object class invariant is of the essence in establishing the correct object semantics.
A better way to deal with the issue would be to check whether the variables are used by the same clients in the same context. As it requires a global analysis of the code, we considered it was too time consuming and could actually be captured in a better way by tracking user direct use of a renaming IDE functionality.
   \item An attribute of the new version that does not appear in the old version generates an ``Attribute added'' SMO. The framework always detects this SMO, initializes the attribute to its default and suggests to check the new class invariant with respect to the old one to avoid deserialization failures.   
   \item An attribute \textit{att} removed from the old version generates an ``Attribute removed'' SMO if in the new version there is no new attribute \textit{att'} having the same type as \textit{att}. The framework always detects this SMO. Once again, the class invariant of the retrieving object will take care of validating the new version semantics. 
   \item An attribute of the old version that has been augmented with the keyword \textit{attached} in the new version generates an ``Attribute attached added"  SMO. The implementation of this SMO into the framework is ongoing work. The idea is to make sure the attribute is attached by explicitly creating an object in the transformation function body, attaching it to the attribute and coding a postcondition stating that the attribute should be non-void. Here both the transformation function postcondition and the retrieving object class invariant are crucial to validate possible semantics changes. 
\end{enumerate}  
To detect the SMOs, we iterate through the new class attributes, search for a match with attributes in the old class and create a corresponding heuristic. We then repeat the process starting from the old class, in order to gather more information, for example to find all the attributes that were removed.     

\subsection{P-evolution-robustness: a measure for object-oriented software evolution robustness}
It can be helpful to statically measure a software system with respect to its robustness to software evolution. Consider a class with  \texttt{m} versions. Given that set of versions, there are \texttt{m(m-1)} possible pairs of versions, each pair representing a transformation function from a version $v_i$ to a version $v_j$ ($i,j=1,\ldots,m, i\ne j$). We define a measure of class p-evolution-robustness (PER) as the ratio of the transitive closure of the binary relation ``there is a transformation function between versions $v_i$ and $v_j$'' to the total number of transformation functions noted above. For instance, considering the two versions of class \textit{BANK\_ACCOUNT} introduced in Section~\ref{sec:motivation}, if someone defines only the forward transformation from version 1 to version 2, their PER would evaluate to 0.5: one defined transformation out of a possible two (one in each direction). As another example, consider to have a class with 5 versions and 5 transformation functions defined, going from version $v_i$ to version $v_{i+1}$ ($i=1,\ldots,4$). Because of the transitive closure, there will be 5 more transformation functions assumed, because the existence of transformation functions between $v_i$ and $v_{i+1}$ and between $v_{i+1}$ and $v_{i+2}$ implies the one between $v_i$ and $v_{i+2}$. Ten transformation functions are therefore assumed (out of 20), giving a PER of 0.5. 

The PER measure can be easily restricted to a single class version $v_i$ by computing the ratio of transformation functions defined between $v_i$ and any other version to the total number of transformation functions that in this case is \texttt{2(m-1)}. 

The PER measure can also be extended to a system release, by computing the average of the PERs of the classes in the release.

In Section~\ref{sec:dataAnalysis} we show how we compute PER for a software system that does not implement the multi-version approach, namely the \texttt{java.util} package and its classes.
\section{Empirical Evaluation: the ESCHER Model vs. the Practice of Object-Oriented Schema Evolution}\label{sec:dataAnalysis}
To assess the relevance of our model, we analyze four sets of classes in two programming languages, Java and Eiffel. We focus on persistence-related code changes. For each language, we choose one set of classes from a widely used data structures and utility library, and one other from a well-known software project. Therefore our investigation takes into account both libraries and applications. The tables shown capture the essential data. The complete set of data is available for download.\footnote{http://tinyurl.com/ESCHER-data (last visited: 15/6/12)}
 
\subsection {Persistence-related code changes in Java}
Advani et al. \cite{AdvaniHassounCounsell06} already evaluated code changes in fifteen open source Java systems and showed that ``rename field'',  ``move field'', ``rename method'', and ``move method'', account for approximately $66\%$ of the total identified code changes. We can additionally observe that ``rename field'' and ``move field'' alone account for $32\%$ of the total. While these results are calculated on all classes of the considered systems, there is no evidence that classes meant to be persistent would exhibit the same characteristics. It seems then reasonable to conjecture that class schema evolution in object-oriented applications is a relevant issue. In particular, the fact that class schema evolution involves attributes in one out of three cases suggests that providing support to handle the issue is a reasonable aim. To improve on the previous study, we focus on classes intended to produce objects that will be persisted. We therefore study a subset of likely-to-be-persisted classes from the Java package \texttt{java.util} to check if they evolve in a similar manner to non-persistent ones.
Java code was a conscious choice because it is widely used and because classes that might have persistent instances are easily recognizable, as they are marked as \texttt{Serializable}. 

\subsubsection {Analyzing \texttt{java.util}}
The package contains classes modeling collections, dates, currencies, and locales. We considered the 22 classes in the package directly implementing the \textit{Serializable} interface and analyzed them manually across five versions of the language: 1.2.2, 1.3.1, 1.4.2, 5.0, 6.0. We chose these versions because they can be considered significant steps in the evolution of the Java language, bearing important changes in both the APIs and their implementations.\footnote{http://en.wikipedia.org/wiki/Java\_version\_history (last visited: 15/6/12)} We also took into consideration 22 code changes types, seven of which are directly relevant to the serialization process and are shown in Table~\ref{tab:view1}. ``Attribute initialization value changed'' is considered because it may influence the class invariant. ``Attribute becoming a constant'' is relevant because constants are not serialized, and therefore it will be interpreted as ``Attribute removed''. In a similar fashion, ``Constant becoming an attribute'' will be interpreted as ``Attribute added''. In line with what was discovered previously, Table~\ref{tab:view1} shows that the persistence-related code changes consist of $17.7\%$ of the total (115 out of 649). Note also that ``Attribute added'' and  ``Attribute removed'' together constitute 74\% of all persistence-related code changes (85 out of 115).

\begin{table}[ht]
\hspace{-3mm}\scriptsize
\caption{Changes found across 5 versions of \texttt{java.util}.}
\label{tab:view1}
\begin{tabular}{l|c|c|c|c|c|c|c|c|c}
&\begin{sideways}attribute added\end{sideways}
&\begin{sideways} attribute removed \end{sideways}
&\begin{sideways} attribute renamed \end{sideways}
&\begin{sideways} attribute type changed \end{sideways}
&\begin{sideways} attribute value changed \end{sideways}
&\begin{sideways} attribute becoming constant \end{sideways}
&\begin{sideways} constant becoming attribute \end{sideways}
&\begin{sideways} non persistence-related \end{sideways}
&\begin{sideways} all\end{sideways}\\
\hline
1.2.2-1.3.1&		4&	0&	0& 0& 1& 0& 0& 11& 16\\	
\hline
1.3.1-1.4.2&		10&	13&	1& 1& 0& 0& 1& 99& 125\\	
\hline
1.4.2-5.0&			29&	6&	0& 13& 2& 4& 0& 270& 324\\	
\hline
5.0-6.0&	    			14&	9&	4& 1& 0& 2& 0& 154& 184\\
\hline
\hline
All&					57& 28& 5& 15& 3&	6&	1& 534& 649\\
\end{tabular}
\end{table}

Table~\ref{tab:view3} shows the distribution of code changes across different classes and across all five versions of the \texttt{java.util} package. The data suggest that persistence-related code changes are sufficiently widespread among classes. 

\begin{table}[!ht]
\caption{Changes found in 5 versions of \texttt{java.util}, by class.}
 \label{tab:view3}
\centering
 \begin{tabular}{l|c|c}
 Class&Code changes&Persistence-related\\
 \hline
 ArrayList&18&3\\
 BitSet&42&8\\
 Calendar&52&24\\
 Currency&3&2\\
 Date&22&9\\
 EnumMap&1&0\\
 EnumSet&1&0\\
 EventObject&1&1\\
 HashMap&101&11\\
 HashSet&9&2\\
 HashTable&41&5\\
 IdentityHashMap&20&2\\
 LinkedHashSet&5&1\\
 LinkedList&50&1\\
 Locale&34&16\\
 PriorityQueue&17&0\\
 Random&13&7\\
 TimeZone&28&7\\
 TreeMap&122&13\\
 TreeSet&39&3\\
 UUID&0&0\\
 Vector&30&0\\
  \hline
   \hline
 All&649&115\\
 \end{tabular}
\end{table}
This study shows that classes whose instances might be serialized change over time.   
Moreover $17.7\%$ of these changes directly impact the capability of classes to deserialize instances of their previous versions. This is aggravated by the fact that there may be a high number of stored objects that need to be updated. \\
In class \textit{java.util.Calendar}, version 1.2, we found two version-related attributes, \textit{currentSerialVersion} and \textit{serialVersionOnStream}, in addition to the usual \textit{serialVersionUID}. As explained in the code comments, these were used to solve the problem of keeping track of the different object versions, and had predefined values suggested. By using ESCHER and its integrated version support, creating these two attributes would have proved unnecessary and the transformation function would have not been cluttering the class itself. In conclusion, while we cannot be sure that deserializing will be performed in a semantically inconsistent way, it is likely that it will create issues at some point.

\subsubsection {The p-evolution-robustness of \texttt{java.util}}
In spite of the fact that Java implements a two-version approach (as seen in Section~\ref{sec:earlierWork}), we are able to compute the PER for the \texttt{java.util} package. While the detailed data are available for download,\footnote{http://tinyurl.com/ESCHER-data (last visited: 15/6/12)} in Table~\ref{tab:pev1} we show how we performed the calculation for class \texttt{ArrayList}. Between versions 1.2.2 and 1.3.1 there are no changes, so both the forward and the backwards transformations work. From version 1.2.2 to 1.4.2 there is an attribute added, for which only the forward transformation works. The reverse transformation is not handled, and so deserialization of objects of version 1.4.2 from version 1.2.2 does not work, because removing an attribute is an incompatible change according to the Java Serialization Specification~\cite{javaspecs}. In a similar way, and taking into account that an attribute type changed is also an incompatible change, we obtain for \texttt{ArrayList} a total of 4 existing transformations, out of 20 possible across the 5 considered versions. The resulting PER value is therefore 0.2. In Table~\ref{tab:pev2} we show the value of PER for every class in the \texttt{java.util}, and for the package itself. 

\begin{table}[!ht]
\caption{P-evolution-robustness (PER) of \texttt{ArrayList}. }
 \label{tab:pev1}
\centering
 \begin{tabular}{l|c|c|l}
 Version interval&Fw. tr.&Bw. tr.&Attribute changes\\
 \hline
 1.2.2-1.3.1&1&1&no changes\\
 1.2.2-1.4.2&1&0&1 added\\
 1.2.2-1.5.0&0&0&1 added, 1 type changed\\
 1.2.2-6.0&0&0&1 added, 2 type changed\\
 1.3.1-1.4.2&1&0&1 added\\
 1.3.1-1.5.0&0&0&1 added, 1 type changed\\
 1.3.1-6.0&0&0&1 added, 2 type changed\\
 1.4.2-5.0&0&0&1 type changed\\
 1.4.2-6.0&0&0&2 type changed\\
 5.0-6.0&0&0&1 type changed\\  
 \hline
  \hline 
 &3&1&PER: 0.2\\
 \end{tabular}
\end{table}

\begin{table}[!ht]
\caption{P-evolution-robustness (PER) of \texttt{java.util}. }
 \label{tab:pev2}
\centering
 \begin{tabular}{l|c}
 Class&Class PER\\
 \hline
 ArrayList&0.20\\
 BitSet&0.40\\
 Calendar&0.25\\
 Currency&0.70\\
 Date&0.30\\
 EnumMap&1.00\\
 EnumSet&1.00\\
 EventObject&0.70\\
 HashMap&0.10\\
 HashSet&0.30\\
 HashTable&0.30\\
 IdentityHashMap&0.40\\
 LinkedHashSet&0.40\\
 LinkedList&0.40\\
 Locale&0.20\\
 PriorityQueue&1.00\\
 Random&0.15\\
 TimeZone&0.30\\
 TreeMap&0.20\\
 TreeSet&0.35\\
 UUID&1.00\\
 Vector&1.00\\
  \hline
  \hline 
 Package PER&0.48\\
 \end{tabular}
\end{table}

The data suggest that the PER measure can be a simple way to highlight an often neglected aspect of software development, and can be easily used to assess the quality of a certain software repository and compare it to others. 

\subsubsection {Analyzing the Apache Tomcat code base}
We choose to analyze Apache Tomcat as an example of a well-known and widely used application. A web server typically needs to store information about shared values that need to survive the stateless HTTP protocol interaction. The Tomcat open source project was initially released in 1999. It includes Catalina as a servlet container, Coyote as HTTP connector and Jasper as JSP engine. We analyze the evolution of persistent classes taking into account the earliest and the latest releases available in the open source repository at the time of writing (3.0 and 7.0 respectively).\footnote{http://svn.apache.org/viewvc/tomcat/ (last visited: 15/6/12)} In Table~\ref{tab:view6} we show, for brevity, the classes in Tomcat's code base in which we actually found  at least one SMO. This happened in 12 out of the 27 classes analyzed. The data confirm what we found in the case of the \texttt{java.util} library classes. We did not notice significantly more changes as we might have expected from an application as opposed to a library project. This in our opinion might be because every software project values stability and backward compatibility, and tries to limit changes to the bare minimum.

\begin{table}[ht]
\hspace{-3mm}\scriptsize
\caption{SMOs found in Tomcat's repository.}
\label{tab:view6}
\begin{tabular}{l|c|c|c|c|c}
&\begin{sideways}attribute added\end{sideways}
&\begin{sideways} attribute removed \end{sideways}
&\begin{sideways} attribute renamed \end{sideways}
&\begin{sideways} attribute type changed \end{sideways}
&\begin{sideways} all\end{sideways}\\
\hline
ValueReference&				0& 2&	 0& 0& 2\\	
\hline
ApplicationParameter&		0&	0&	0& 1& 1\\	
\hline
FilterDef&						1&	0&	0& 0& 1\\
\hline
FilterMap&						2&	0&	2& 2& 6\\
\hline
MultipartDef&					0&	0&	0& 3& 3\\
\hline
NamingResources&			2&	0&	0& 0& 2\\
\hline
SecurityCollection&			2&	0&	0& 0& 2\\
\hline
ServletDef&						4&	0&	0& 0& 4\\
\hline
CsrPreventionFilter&			4&	0&	0& 0& 4\\
\hline
AbstractReplicatedMap&		2&	0&	0& 0& 2\\
\hline
FeatureInfo&					2&	1&	0& 0& 3\\
\hline
DeltaSession&					0&	1&	0& 0& 1\\
\hline
 \hline
All&								19& 4& 2& 6& 31\\
\end{tabular}
\end{table}

 \subsection{Persistence-related code changes in Eiffel}

\subsubsection{Eiffel libraries code changes}
We analyze 22 Eiffel classes, most of them direct counterparts of the ones previously analyzed in Java. In the five cases in which it is not possible to find a direct counterpart, we choose classes whose instances are likely to be serialized, like strings and data structures. We also analyze five releases, but the time period on which we focus is wider, ranging from 1993 to 2010 instead that from 1999 to 2009. The main reason is that given that Eiffel is less widespread than Java, we hope to find more interesting data in a wider time interval. In both cases we allow both languages to have an initial stabilization period. The availability of data in the software repositories is also a limiting factor.   
The list of relevant code changes for Eiffel persistence allows us to extend the previous list by adding ``inheritance from one class added'', ``inheritance from one class removed'', ``class name changed'', and ``Void safety clause added''.
With respect to the Java study, we do not consider "Attribute initialization value changed" because in Eiffel assigning a value to an attribute in the declaration means making it a constant, which is typically not serialized. Furthermore, we have not found any instance of "Attribute renamed", "Attribute becoming a constant" or "Constant becoming an attribute". Interestingly, however, we find a consistent number of inheritance-related code changes, of which there was no trace in Java. This may be due to the fact that the Eiffel object model allows multiple inheritance, making it easier to add and remove functionality by using inheritance itself. As we model inheritance by considering the flattened class schema, the "Inheritance added" and "Inheritance removed" code changes fall back into the other already analyzed ones. Finally, there are also some changes to attributes meant to enforce void-safety, a matter discussed in a separate paragraph below. The kinds of code changes found, together with their occurrences and across considered versions, are synthesized in Table~\ref{tab:view4}. Note that ``Attribute added'' and  ``Attribute removed'' are still relevant, while ``Inheritance added" and ``Inheritance removed" together constitute 54\% of all persistence-related code changes.  
\begin{table}[ht]
\hspace{-3mm}\scriptsize
\caption{SMOs found across 5 versions of Eiffel.}
\label{tab:view4}
\begin{tabular}{l|c|c|c|c|c|c|c|c}
&\begin{sideways}attribute added\end{sideways}
&\begin{sideways} attribute removed \end{sideways}
&\begin{sideways} attribute type changed \end{sideways}
&\begin{sideways} inheritance added\end{sideways}
&\begin{sideways} inheritance removed\end{sideways}
&\begin{sideways} class name changed\end{sideways}
&\begin{sideways} void safety added\end{sideways}
&\begin{sideways} all\end{sideways}\\
\hline
1993-2002&		8&	2&	0& 8& 14& 0& 0& 32\\	
\hline
2002-2006&		2&	0&	3& 4& 0& 0& 0& 9\\	
\hline
2006-2008&		4&	1&	0& 3& 0& 1& 0& 9\\	
\hline
2008-2010&	    	0&	0&	0& 0& 0& 0& 4& 4\\
\hline
\hline
All&				14& 3& 3& 15& 14&	1&	4& 54\\
\end{tabular}
\end{table}

Table~\ref{tab:view5} shows the distribution of code changes across different classes and across all the considered five versions of Eiffel. 
\begin{table}[!ht]
\caption{SMOs found across 5 Eiffel versions, per class.}
 \label{tab:view5}
\centering
 \begin{tabular}{l|c}
 Class&Persistence-related changes\\
 \hline
 Arrayed\_list&2\\
 Bit\_ref&0\\
 Date\_time\_parser&6\\
 Array&1\\
 Date&2\\
 String\_8&6\\
 Tuple&3\\
 Action\_sequence&2\\
 Array\_2&0\\
 Arrayed\_set&0\\
 Hash\_table&12\\
 Linked\_tree&1\\
 Linked\_set&2\\
 Linked\_list&0\\
 I18n\_locale\_manager&4\\
 Priority\_queue&0\\
 Random&3\\
 Time&2\\
 Binary\_search\_tree&3\\
 Binary\_search\_tree\_set&4\\
 Uuid&1\\
 Arrayed\_stack&0\\
  \hline
  \hline
 All&54\\
 \end{tabular}
\end{table}
As in the Java study, the data suggest that persistence-related code changes are sufficiently widespread among classes. \\
During our analysis we found a particularly interesting example of the kind of issues ESCHER would help solving.
In class \textit{HASH\_TABLE}, considered from revision 13752 to revision 47039, there was a switch to a more efficient container by changing the types of three attributes. This resulted in the previously serialized objects not being readable anymore. The solution, implemented in a later version, was to code a transformation function using an ad hoc created attribute, \textit{hash\_table\_version\_57}. If the library maintainers had used ESCHER, there would have been no need to add any attribute, and the transformation function would not have been cluttering the class.

\textbf{Void safety-related code changes.}
When we refer to void-safety we mean the guarantee that, in an object-oriented program, the operation x.f (...) will never fail because of the target x denoting a void reference at execution time \cite{Meyer05}. The Eiffel language incorporates the mechanism into the compiler, requiring developers to choose, for every attribute, one keyword as a default (\textit{attached} or \textit{detachable}) and to specify the other when needed. In case of the class \textit{DATE\_TIME\_PARSER} we have discovered four code changes related to void safety, in which the keyword \textit{detachable} was added to four attributes. This code change (from attached to detachable) is harmless under the point of view of software evolution, because an attached attribute, which is forced to be non-void, will always be stored as such and so it will comply with the requirement of being detachable at retrieval time. Though it did not occur in our findings, the symmetric code change, from detachable to attached, can in principle be dangerous. A detachable attribute stored as void will be considered as attached by the retrieving system. This will raise an exception at runtime, both in case of retrieval from storage and in any other case involving assignment at runtime, such as a call to an external module assigning void to an attached attribute. We believe that this finding, while related to the Eiffel language, can still be useful for future evolution of languages like Java that have not yet incorporated this feature.

\subsubsection {Analyzing the EiffelStudio code base}
We choose to analyze EiffelStudio as an example of an IDE, a very common type of application that is likely to have persistence needs. EiffelStudio stores information about user preferences, and also stores data concerning automatically generated tests. The EiffelStudio IDE was initially called EiffelBench and was released in 1990. It was renamed to EiffelStudio in 2001 and was open sourced in 2006. We analyze the evolution of persistent classes of the main EiffelStudio application code and of its automated testing facility called "AutoTest".  We take into account the earliest and the latest releases available in the open source repository at the time of writing (5.4 from 2004 and 6.8 from 2011).\footnote{https://svn.eiffel.com/eiffelstudio/trunk/ (last visited: 15/6/12)} In Table~\ref{tab:view7} we show, for brevity, the classes in which we actually found at least one SMO. This occurred in 12 out of the 20 classes analyzed. The data once again confirm what we found in the case of the Eiffel library classes, and are also consistent with what we found in the Java repositories.

\begin{table}[ht]
\hspace{-3mm}\scriptsize
\caption{SMOs found in EiffelStudio's repository.}
\label{tab:view7}
\begin{tabular}{l|c|c|c|c|c|c|c|c}
&\begin{sideways} attribute added\end{sideways}
&\begin{sideways} attribute removed \end{sideways}
&\begin{sideways} attribute renamed \end{sideways}
&\begin{sideways} attribute type changed \end{sideways}
&\begin{sideways} inheritance added \end{sideways}
&\begin{sideways} class name changed \end{sideways}
&\begin{sideways} void safety added \end{sideways}
&\begin{sideways} all\end{sideways}\\
\hline
Sd\_config\_data&					8& 0& 1& 0& 0& 0& 0& 9\\	
\hline
Search\_table&						2& 0&	 0& 0& 0& 0& 1& 3\\
\hline
Sd\_inner\_container\_data	&			0& 0&	 0& 2& 0& 0& 4& 6\\
\hline
Ev\_split\_area&						1& 0&	 0& 0& 0& 0& 2& 3\\
\hline
Ev\_split\_area\_i&					0& 1&	 0& 0& 0& 0& 3& 4\\
\hline
Ev\_widget\_i&							1& 0&	 0& 0& 0& 0& 3& 4\\
\hline
Sd\_place\_holder\_zone&				1& 0&	 0& 0& 0& 0& 0& 1\\
\hline
Sd\_notebook\_upper&			4& 3&	 0& 2& 0& 0& 0& 9\\
\hline
Epa\_test\_case\_info&			0& 0&	 0& 0& 0& 1& 0& 1\\
\hline
Profile\_information&			0& 0&	 0& 1& 0& 0& 0& 1\\
\hline
Profile\_set&							0& 0&	 0& 36& 0& 0& 0& 36\\
\hline
Test\_session\_record&			0& 0&	 0& 0& 1& 0& 0& 1\\
\hline
\hline
All&											17& 4& 1& 41& 1& 1& 13& 78\\

\end{tabular}
\end{table}
\section{Implementation}\label{sec:implementation}
The ESCHER tool implements the model presented in Section~\ref{sec:model} making it usable in practice. It is integrated into EiffelStudio, and can also be used independently as a library.
While our approach can be applied to any object-oriented programming language providing support for storing and retrieving objects, we use Eiffel (and its IDE EiffelStudio) because of its integrated support for Design by Contract, and in particular for class invariants~\cite{ecma367}. 
 
The detailed documentation about ESCHER, a step-by-step illustrated tutorial, the source code and the executables for the tool are available for download on the ESCHER project page.\footnote{https://trac.inf.ethz.ch/trac/meyer/eve/wiki (last visited: 15/6/12)}

\subsection{Version handling}
ESCHER handles class versions either by using the existing version information, or by automatically assigning a version tag if not present. The versioning mechanism relies on the notion of \textit{release}: a versioned, compiled and semantically coherent set of classes constituting a software system. Different versions of the same class can only be part of different releases. At release time, ESCHER automatically updates the class version tags of the classes that actually changed. The release version is increased if at least one class has changed with respect to the previous release.  
Provided the required transformation functions exist, it will always be possible to attempt a retrieval of any object of a certain class and version into an object of another version of the same class. This applies to both forwards and backwards transformations, as customers might be running an old system (specified by an old release number) and in need of retrieving objects stored by a newer system release. 
\subsection{Code generation}
While modest, the support for generating code provided by ESCHER relieves developers from writing boilerplate code to transform instances stored in different versions with respect to the current one. It also lets developers focus on the actual specification of the transformation function rather than on the framework code necessary to wire all the transformation functions. Given a class  \texttt{C}, ESCHER instruments it with a version tag, and automatically creates a corresponding handler class containing all the related transformation functions. The generated name of the handler class is \texttt{C\_SCHEMA\_EVOLUTION\_HANDLER} in this case. All the associations between each handler name and its corresponding class are kept in the ad hoc generated class \texttt{SCHEMA\_EVOLUTION\_PROJECT\_MANAGER}. ESCHER automatically creates and manages all the separate folders containing the various releases as well. The code generation algorithm for the transformation functions compares the two abstract syntax trees of two class versions statically, looking for known SMOs. If a known SMO is detected, the algorithm generates a transformation function and code according to the respective heuristic rule. The transformation function is then placed into an ad hoc handler class, 
containing all the existing transformation functions between pairs of different versions. This helps to avoid polluting the class code with evolution-related code, and makes the whole approach more scalable. Comments, hints and informational messages are generated to guide developers through the process. 
\subsection{IDE support and runtime mechanism}
ESCHER is seamlessly integrated into the EiffelStudio IDE, and its functionalities are accessible through an ad hoc panel. In addition to release a project, developers can trigger the creation and release of a transformation function by selecting two versions from a dynamically generated panel that only lists the versions actually existing in the repository. Developers can also create a ``filter'', that allows them to select the attributes they want to be part of the serialized form.

At retrieval time, ESCHER relies on an ad hoc algorithm automatically performing conversions across two different versions. The algorithm will raise an exception either when a specific schema evolution handler does not exist, or when a specific transformation function between the two versions does not exist. The class invariant of the retrieving object is also enforced. 
\section{Assessment and Future Work}\label{sec:assessment_future_work}

 \subsection{Evaluating ESCHER on the Java and Eiffel repositories}
Analyzing the history of classes along different versions provides useful insights on the schema evolution process. The data analysis presented in Section~\ref{sec:dataAnalysis} shows that the schema of persistent classes may evolve significantly both in software libraries and in applications. To evaluate ESCHER, we check if it recognizes the 278 persistence-related SMOs (counting both Eiffel and Java SMOs) found in our analysis. The results are that ESCHER is able to identify the SMOs in 93\% of all cases (259 out of 278). The unrecognized SMOs are "Void safety added" and "Class name changed", which is interpreted as a new class introduced into the system. We count "Attribute renamed" as recognized because ESCHER, though not generating the full body of the transformation function, it is aware of the different possibilities and generates a comment providing guidance to developers. We also observe that in none of the four code repositories analyzed an "Attribute renamed" was a false positive, that is, had to be interpreted as an "Attribute removed" plus "Attribute added". The results above suggest that the simple model we devised is sufficiently detailed to take into account a realistic set of SMOs in two different object-oriented programming languages.

We have shown in Section~\ref{sec:model} that the transformations functions are complete with respect to attribute modifications. With respect to their soundness, we do not provide static guarantees, but given any two versions, and given a transformation function between the two, the tool ensures that the function pre- and post-conditions, and the retrieving class invariant, hold at runtime.

We have already remarked the importance of class invariants to avoid that potentially unsafe objects are accepted into the system. Unfortunately, class invariants are not yet widely accepted among developers, and not all mainstream languages offer native support for them. An obvious question is then: what can Java developers do to emulate the ESCHER mechanism enforcing class invariants? While analyzing the Java libraries, we discovered  a possible solution in class \textit{BitSet}, during the transition from version 5.0 to 6.0. A new method \textit{checkInvariants()} is introduced, with the idea of enforcing some class-wide properties. This method is then invoked from every public method in the class, emulating the Eiffel invariant checking mechanism. In conclusion, the absence of an embedded Design By Contract language support (also for preconditions and postconditions) makes things more complicated for Java developers, but the example of class \textit{BitSet} shows that emulating the invariant mechanism remains possible. 

In the worst case, developers can still bypass, actively or passively, all the checks mentioned. In case of an attribute added, for example, they could ignore the warning, accept the (possibly wrong) default provided, and in general fail to code an appropriate transformation function or class invariant. They could even code the right invariant but disable runtime checks for assertions. This means that while ESCHER supports developers in handling the class schema evolution of an object-oriented software application, it does not guarantee a completely safe class schema evolution. To improve this, we believe that our measure for evolution robustness can help in assessing how much trust we can put in a code release with respect to class schema evolution. 

 \subsection{Threats to validity}
While the analysis of the results of our four studies shows that the ESCHER model and its implementation address most practical challenges, the validation based on empirical data is by definition questionable. In addition, the current SMO detection and code generation could be dependent on the specific programming languages chosen (Java and Eiffel). 

Assuming that the idea and the current implementation are meaningful, general willingness to adopt ESCHER remains to be proved. In particular, there is no usability study trying to investigate developers' understanding and acceptance of such a solution. We did not perform a usability study because we gave priority to validating ESCHER using existing, realistic and widely known software projects. 

 \subsection{Conclusions}
This work proposes a shift of attitude with respect to how developers cope with class schema evolution. By integrating support into an IDE, the approach elevates class schema evolution to the status of first-class citizen of the software development process rather than undesirable side effect of the software production activities. It also proposes a significant time shift (from runtime to release time) for the evolution-related activities. A system release becomes then an event triggering a whole set of tool actions intended to help developers to better focus on possible issues that may arise from previous versions of the newly released code.

Possible future work is to broaden the empirical validation by performing studies on more code repositories, in particular in object-oriented programming languages other than Java and Eiffel. We also plan to integrate object browsing and writing-back capabilities into the IDE to help developers having to analyze objects previously serialized using a non-human-readable format. Finally we plan to consider, in our model of software evolution, the evolution of the persistence software mechanisms themselves.

\ifCLASSOPTIONcompsoc
  \section*{Acknowledgments}
\else
  \section*{Acknowledgment}
\fi

The authors thank the Swiss National Foundation (grant number 200021-113808) for partially funding this work. They also thank Matthias Loeu and Teseo Schneider for their implementation work on the initial versions of ESCHER.

\ifCLASSOPTIONcaptionsoff
  \newpage
\fi



%

\bibliographystyle{IEEEtran}
\bibliography{./IEEEabrv,./picmarcobiblio}

\begin{IEEEbiographynophoto}{Marco Piccioni}
Marco Piccioni is research assistant and PHD student at the Chair of Software Engineering at ETH Zurich. After a degree in Mathematics at Universit\`{a} degli Studi di Roma La Sapienza, and a master in Economics at Universit\`{a} L. Bocconi, Milano, he worked for ten years at Sistemi Informativi S.p.A. (an IBM company) as a software engineer and trainer. His research work at ETH focuses on the software evolution of persistent object-oriented applications.
\end{IEEEbiographynophoto}


\begin{IEEEbiographynophoto}{Manuel Oriol}
Manuel Oriol is a Principal Scientist with ABB Corporate Research, Industrial Software Systems in Baden-D\"attwil, Switzerland. He is also a Lecturer at University of York in the United Kingdom. Manuel graduated from ENSEEIHT school with an Engineer diploma in Computer Science and an MSc in Artificial Intelligence from INP Toulouse. He subsequently graduated from University of Geneva with a PhD in Information Systems. His research interests span over a large spectrum and include object-oriented languages, software testing, component infrastructure, computer science education, middleware/cloud, real-time systems, and dynamic software updating.

\end{IEEEbiographynophoto}

\begin{IEEEbiographynophoto}{Bertrand Meyer}
Bertrand Meyer is Professor of Software Engineering at ETH Zurich, head of the Software Engineering Laboratory at ITMO (Saint Petersburg, Russia), and Chief Architect of Eiffel Software.
\end{IEEEbiographynophoto}




\end{document}